\documentclass[12pt]{article}

\usepackage{amsmath,amssymb}

\usepackage[usenames]{color}

\newcommand{\beq}{\begin{eqnarray}}
\newcommand{\eeq}{\end{eqnarray}}

\begin{document}

%\begin{widetext}

\thispagestyle{empty}

\begin{center}

\begin{flushright}
BCCUNY-HEP/07-05
\end{flushright}

\vspace{15pt}
{\large \bf COMPOSITE STRINGS
IN $(2+1)$-DIMENSIONAL 
ANISOTROPIC WEAKLY-COUPLED YANG-MILLS THEORY  }

%\textcolor{red}{text in red}

\vspace{20pt}

%\textcolor{green}{text in green}

{\bf Peter Orland}$^{\rm a.b.c.d.}$\!\footnote{orland@nbi.dk, 
giantswing@gursey.baruch.cuny.edu}

\vspace{8pt}

\begin{flushleft}
a. The Isaac Newton Institute for the Mathematical Sciences, 20 Clarkson Road,
Cambridge, CB3 OEH, UK
\end{flushleft}
 
\begin{flushleft}
b. The Niels Bohr Institute, The Niels Bohr International Academy, Blegdamsvej 17, DK-2100, Copenhagen {\O}, Denmark
\end{flushleft}
 
\begin{flushleft}
c. Physics Program, The Graduate School and University Center,
The City University of New York, 365 Fifth Avenue,
New York, NY 10016, U.S.A.
\end{flushleft}

\begin{flushleft}
d. Department of Natural Sciences, Baruch College, The 
City University of New York, 17 Lexington Avenue, New 
York, NY 10010, U.S.A. 
\end{flushleft}

\vspace{40pt}

{\bf Abstract}

\end{center}

\noindent 
The small-scale structure of a string connecting a pair of static sources is explored for
the weakly-coupled anisotropic SU($2$) Yang-Mills theory in $(2+1)$ dimensions. A 
crucial ingredient in the formulation of
the string Hamiltonian is the phenomenon of color smearing of 
the string constituents. The quark-anti-quark potential
is determined. We close with some discussion of the standard, fully Lorentz-invariant
Yang-Mills theory.

%\end{widetext}

\hfill

\newpage

\setcounter{footnote}{0}

\setcounter{page}{1}

\section{Introduction}
\setcounter{equation}{0}
\renewcommand{\theequation}{1.\arabic{equation}}

Recently, the author has established the existence of confinement and a mass gap
in a version of $(2+1)$-dimensional SU($N$) Yang-Mills theory, in which the
coupling constants are anisotropic and small. The understanding of
the inter-quark potential and the mass gap is elementary \cite{PhysRevD71}, 
\cite{PhysRevD75-1} though
finding precise values for the string tension  \cite{PhysRevD74}, and the mass spectrum
\cite{PhysRevD75-2} requires detailed information of an integrable $(1+1)$-dimensional
quantum field theory. This integrable field theory is the 
${\rm SU}(N) \times {\rm SU}(N)$ principal-chiral nonlinear
sigma model. For 
$N=2$, exact knowledge of certain matrix elements 
makes it possible to perturb away from integrability.

Though the gauge theory
we consider is not spatially-rotation invariant, it 
has features one expects of real $(3+1)$-dimensional QCD; it is
asymptotically free and confines quarks at weak coupling. 

One can formally remove the regulator in strong-coupling
expansions of $(2+1)$-dimensional gauge theories; the vacuum state in
this expansion yields a string tension and a mass gap which have
formal continuum limits. This is possible because
of purely dimensional considerations in this number of
dimensions. Such strong-coupling analyses  
can be done in a Hamiltonian lattice formalism \cite{Greensite}, 
or with an ingenious choice of degrees of freedom and point-splitting regularization 
\cite{kar-nair}. There are even 
formal improvements of the vacuum
state using the point-splitting cut-off \cite{ag-kar-nair} 
or the lattice cut-off \cite{green-olej}, which do not confine
adjoint sources. It is important to know whether these results
can be justifiably extrapolated to the limit of no
regularization
(more discussion of this issue can be found in 
the introduction of reference
\cite{PhysRevD74}). In contrast, the approach we have taken
is a weak-coupling method. It is, thus far, the only method
yielding quark confinement with no strong-coupling assumptions 
in more than two dimensions, without
dynamical matter. There is a hint of another
weak-coupling approach in $(2+1)$ dimensions \cite{kar-nair2}, \cite{orl-sem}
based on general properties of gauge-orbit space.

Simple intuitive formulas for the potential between a static quark and antiquark
were found quite early for our anisotropic theory
\cite{PhysRevD71}. String tensions for higher
representations can also be worked out, and adjoint sources are not
confined  \cite{PhysRevD75-1}. The string tensions for the cases of {\em horizontally}
and {\em vertically} separated quarks, {\em i.e.}  separated in the $x^{1}$-
and $x^{2}$-directions, respectively, have corrections, however. For gauge group SU($2$), the 
leading
correction to the horizontal string tension was found in reference \cite{PhysRevD74}. 
In 
this paper, the vertical potential is shown to be the ground-state energy
of a certain Hamiltonian in one spatial dimension. This Hamiltonian describes
the dynamics of a string with both
coordinate and color degrees of freedom. The correction to the potential of a vertically
separated quark-antiquark pair is thereby determined.

The connection between the gauge theory and integrable systems
using the Kogut-Susskind lattice formalism was
explained in references \cite{PhysRevD71}, \cite{PhysRevD74}. A quicker derivation 
was given in references \cite{Conf7}, \cite{PhysRevD75-2}. Here we simply present the axial-gauge
Hamiltonian formalism and refer the reader
to these papers for its derivation.

The $2$-coordinate is discrete, so that $x^{2}$ takes the
values $x^{2}=a,2a,3a\dots, L^{2}$, where
$a$ is a lattice spacing. All 
fields are considered functions of $x=(x^{0},x^{1}, x^{2})$.  The boundary condition 
is periodic in $x^{2}$, which means that any function
$f(x^{0},x^{1},x^{2})$ satisfies $f(x^{0},x^{1},x^{2}+L^{2})=f(x^{0},x^{1},x^{2})$. The
boundary condition
in the $x^{1}$-direction is open, so that
space is a cylinder. In this paper, we assume the thermodynamic
limit is already taken, so we will not worry too much about the boundaries. The
gauge fields are SU($N$)-Lie-algebra valued. We chose generators
of this Lie algebra $t_{b}$, which satisfy ${\rm Tr}\,\,t_{b}t_{c}=\delta_{bc}$ and
define structure coefficients $f^{d}_{bc}$ by
$[t_{b},t_{c}]={\rm i}f^{d}_{bc}t_{d}$.  We have set the gauge component $A_{1}(x)$ to be zero and 
replace $A_{2}(x)$ by a field $U(x)$ lying in
SU($N$), via 
\beq
U(x)= \exp i\int_{x^{2}}^{x^{2}+a} dy^{2}A_{2}(x^{0},x^{1},y^{2})\;.
\nonumber
\eeq 
The left-handed and right-handed currents are, 
\beq
j^{\rm L}_{\mu}(x)_{b}={\rm i}{\rm Tr}\,t_{b} \, \partial_{\mu}U(x)\, U(x)^{\dagger}\;,\;\;
j^{\rm R}_{\mu}(x)_{b}={\rm i}{\rm Tr}\,t_{b} \, U(x)^{\dagger}\partial_{\mu}U(x)\;,
\nonumber 
\eeq 
respectively, 
where $\mu=0,1$. The Hamiltonian is $H_{0}+H_{1}$, where
\beq
H_{0}\!=\!\sum_{x^{2}}\int dx^{1} \frac{1}{2g_{0}^{2}}\{ [j^{\rm L}_{0}(x)_{b}]^{2}+[j^{\rm L}_{1}(x)_{b}]^{2}\}
\;,\label{HNLSM}
\eeq
and
\beq
\!\!\!&\!\!H_{1}\!\!&\!\! =
-\sum_{x^{2}}  \int \! dx^{1}\!\!\int \!dy^{1}
\frac{(g_{0}^{\prime})^{2}}{4g_{0}^{4}a^{2}}\, \vert x^{1}-y^{1}\vert \nonumber \\
\!\!&\!\!\times\!\!&\!\!\!\!\!
\left[ j^{\rm L}_{0}(x^{1},x^{2})_{b}\!-\!j^{\rm R}_{0}(x^{1},x^{2}-a)_{b}
\!-\!
{\bar {\rm q}}_{b}\delta(x^{1}\!-\!u^{1})
\delta_{x^{2} u^{2}}\!+\!
{\rm q}_{b}\delta(x^{1}\!-\!v^{1})
\delta_{x^{2} v^{2}}
\right]
\nonumber \\
\!\!&\!\!\times\!\!&\!\!\! \!\!
\left[j^{\rm L}_{0}(y^{1},x^{2})_{b}\!-\!j^{\rm R}_{0}(y^{1},x^{2}-a)_{b}
\!-\!
{\bar {\rm q}}_{b}\delta(y^{1}\!-\!u^{1})
\delta_{x^{2} u^{2}}\!+\!
{{\rm q}}_{b}\delta(y^{1}\!-\!v^{1})
\delta_{x^{2} v^{2}}
\right]\!,
\label{continuum-local}
\eeq
where 
we have inserted two color charges - a quark with charge $\rm q$ at site $v$
and an anti-quark with charge ${\bar {\rm q}}$ at site $u$. These charge
operators satisfy $[{\rm q}_{b},{\rm q}_{c}]={\rm i}f_{bc}^{d}{\rm q}_{d}$
and $[{\bar {\rm q}}_{b},{\bar {\rm q}}_{c}]={\rm i}f_{bc}^{d}{\bar {\rm q}}_{d}$. A constraint remains
after the axial-gauge fixing, namely that 
for each $x^{2}$
\beq
\int d x^{1}\left[ j^{L}_{0}(x^{1},x^{2})_{b}-j^{R}_{0}(x^{1},x^{2}-a)_{b}
-{\bar {\rm q}}_{b}\delta(y^{1}\!-\!u^{1})
\delta_{x^{2} u^{2}}\!+\!
{{\rm q}}_{b}\delta(y^{1}\!-\!v^{1})
\delta_{x^{2} v^{2}}\right] 
\Psi=0,
\label{physical}
\eeq
where $\Psi$ is any physical 
state. For more details on the derivation of the term in the Hamiltonian
(\ref{continuum-local}) and the constraint (\ref{physical}), see
references
\cite{PhysRevD71}, \cite{PhysRevD74}. The Hamiltonian $H_{0}$ given in (\ref{HNLSM}) is a sum of principal-chiral sigma-model Hamiltonians.

The anisotropic regime of  $(2+1)$-dimensional Yang-Mills theory is
\beq
(g_{0}^{\prime})^{2} \ll \frac{1}{g_{0}}e^{-4\pi/(g_{0}^{2}N)}\;. \label{relative-scales}
\eeq
The point
where the regulator can be removed in the theory is the same as that of
the standard isotropic theory $g_{0}^{\prime}=g_{0}=0$. The 
left-hand side and ride-hand side are proportional to the two
energy scales in the theory (the latter comes from the two-loop beta function of
the sigma model). For more discussion of these matters, see references \cite{PhysRevD75-1},
\cite{PhysRevD74}  and
\cite{PhysRevD75-2}.

The 
excitations of $H_{0}$, which we call Fadeev-Zamolodchikov or FZ particles, behave like 
solitons, though they are not quantized versions of classical solutions. Some of these FZ particles 
are elementary and others are bound states of
the elementary FZ particles. An elementary FZ particle has an adjoint charge and mass $m_{1}$. An 
elementary one-FZ-particle state
is a superposition of color-dipole states, with a quark  (anti-quark)
charge at $x^{1}, x^{2}$ and an anti-quark (quark)
charge at $x^{1},x^{2}+a$.  The interaction
$H_{1}$ produces a linear potential between color charges with the same value of $x^{2}$. Residual gauge
invariance (\ref{physical}) requires that at each value of $x^{2}$, the total color charge is zero. If there are 
no quarks with coordinate $x^{2}$, the total right-handed charge of FZ particles in the sigma model
at $x^{2}-a$ is equal to the total left-handed charge of FZ particles in the sigma model at $x^{2}$.

The particles of the principal-chiral sigma model carry a quantum number 
$r$, with the values $r=1,\dots,N-1$
\cite{abda-wieg}.  Each particle of label $r$ has an antiparticle 
of the same mass, with label $N-r$. The
masses are given by
\beq
m_{r}=m_{1}\frac{\sin\frac{r\pi}{N}}{\sin\frac{\pi}{N}},\;\; m_{1}=K\Lambda(g_{0}^{2}N)^{-1/2}e^{-\frac{4\pi}{g_{0}^{2}N}}+{\rm non\!-\! universal \;corrections}\;, \label{mass-spectrum}
\eeq
where $K$ is a non-universal constant and $\Lambda$ is the ultraviolet cut-off
of the sigma model.
 
Lorentz invariance
in each $x^{0},x^{1}$ plane is manifest. For this reason, the 
linear potential is not the only effect of
$H_{1}$. The interaction creates and destroys pairs of elementary
FZ particles. This effect is quite small, provided that $g_{0}^{\prime}$ is small 
enough. Specifically, this means that the string tension in the $x^{1}$-direction coming from $H_{1}$ is small compared to
the square of the mass of the fundamental FZ particle; this is just the condition
(\ref{relative-scales}). The effect is important, however, in that it is responsible for the correction to
the horizontal string discussed in the 
next paragraph in equation (\ref{string-tension}).

Simple arguments readily show that at leading order in $g_{0}^{\prime}$, the vertical and
horizontal string tensions are given by
\beq
\sigma_{\rm V}=\frac{m_{1}}{a}\;,\;\; \sigma_{\rm H}=
\frac{(g_{0}^{\prime})^{2}}{a^{2}}C_{N}\;, \label{string-tensions}
\eeq
respectively, where
$C_{N}$ is the smallest eigenvalue of the Casimir of 
${\rm SU}(N)$. 
These naive results for the string tension have further corrections in $g_{0}^{\prime}$, which were 
determined for the horizontal string tension for SU($2$) \cite{PhysRevD74}:
\beq
\sigma_{\rm H} \!\!&\!\!=\!\!&\!\! \frac{3}{2}\frac{(g_{0}^{\prime})^{2}}{a^{2}} \left[ 1+
\frac{4(g_{0}^{\prime})^{2}}{3\pi^{2}m_{1}^{2}a^{2}}
\exp-2\!\int_{0}^{\infty} \frac{d\xi}{\xi}e^{-\xi}\tanh^{2}\frac{\xi}{2}
\right]^{-1} \nonumber \\
\!\!&\!\!=\!\!&\!\!
\frac{3}{2} \frac{(g_{0}^{\prime})^{2}}{a^{2}} \left[ 1+
\frac{4(g_{0}^{\prime})^{2}}{3\pi^{2}m_{1}^{2}a^{2}}\,\,0.7296\right]^{-1}\;.
\label{string-tension}
\eeq
The leading term agrees with (\ref{string-tensions}). This calculation was done using the exact form factor for sigma model currents obtained by
Karowski and Weisz \cite{KarowskiWeisz}. In this paper, we shall use the form factor
to study corrections of order $(g_{0}^{\prime})^{2}$ to the vertical string tension. A review of
integrability and form-factor methods is in the appendix of reference 
\cite{PhysRevD74}.

A picture of a gauge-invariant state for the gauge group SU($2$) with a single
quark and a single antiquark at different values of $x^{2}$
is given in Figure 1. For $N>2$, there
are more complicated ways in which strings can join particles. The lightest states have
the smallest number of particles, by virtue of $\sigma_{\rm H}\ll \sigma_{\rm V}$. Thus, there
is a single FZ particle in each layer between the quark and the antiquark. There is a piece
in $H_{1}$ which can create and destroy FZ particles, but this can safely be neglected
in a nonrelativistic approximation. We shall treat the quarks as static, non-dynamical
sources in this paper.

\begin{center}

\begin{picture}(150,60)(30,0)

\linethickness{0.5mm}

\multiput(75,-0.5)(0,7){8}{\multiput(0,0)(5,0){6}{\put(0,0){$-$}}}
\multiput(75,-0.5)(0,7){8}{\multiput(35,0)(5,0){1}{\put(0,0){$-$}}}
\multiput(75,-0.5)(0,7){8}{\multiput(45,0)(5,0){3}{\put(0,0){$-$}}}

\put(105,-0.5){$-$}
\put(105,6.5){$-$}
\put(105,13.5){$-$}
\put(105,20.5){$-$}
\put(105,27.5){$-$}
\put(105,34.5){$-$}
\put(105,48.5){$-$}

\put(115,-0.5){$-$}
\put(115,13.5){$-$}
\put(115,20.5){$-$}
\put(115,27.5){$-$}
\put(115,34.5){$-$}
\put(115,41.5){$-$}
\put(115,48.5){$-$}

\put(95,11.5){\circle*{7}}
\put(85,18.5){\circle*{7}}
\put(110,25.5){\circle*{7}}
\put(99,32.5){\circle*{7}}
\put(120,39.5){\circle*{7}}

\put(95,8.0){\line(1,0){20}}
\put(85,15.2){\line(1,0){10}}
\put(110,22.0){\line(-1,0){25}}
\put(99,29.0){\line(1,0){10.5}}
\put(120,36.0){\line(-1,0){21}}
\put(110,42.8){\line(1,0){10}}

\put(107,42.7){\circle{5}}
\put(105.7,42){$\bf q$}

\put(118,8.2){\circle{5}}
\put(116.7,7.4){${\bf{\bar {\bf q}}}$}

\end{picture}
\end{center}

\vspace{5pt}

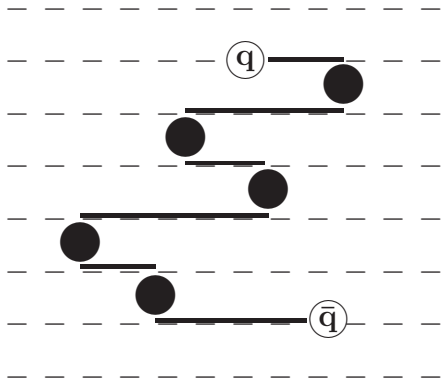
Figure 1. A low-lying quark-antiquark-pair state. The 
horizontal coordinate is $x^{1}$ and the vertical coordinate is $x^{2}$. The quark
lies at a larger value of $x^{2}$ than the antiquark. Between the pair is a
collection of FZ particles. All the particles are bound together by horizontal
strings.

\vspace{15pt}

In the next section we show how the color of FZ particles is smeared by radiative
corrections, with the aid of the exact matrix elements of the current operator. We use
this to derive the Hamiltonian of a string in Section 3. The ground-state energy of this
string, and thus the potential between static color sources is found in Section 4.  In Section
5, we argue that the functional form of this potential extends to the standard
Lorentz-invariant SU($2$) Yang-Mills theory. We 
present our conclusions in Section 6.

\section{Color smearing}
\setcounter{equation}{0}
\renewcommand{\theequation}{2.\arabic{equation}}

Consider a static quark-antiquark pair for the SU($2$) gauge theory, as 
in Figure 1. We will assume that
the $x^{1}$-coordinate of the quark and antiquark is the same and that
the $x^{2}$-coordinate of the quark is $v^{2}$ and the $x^{2}$-coordinate
of the antiquark is $u^{2}$, where $v^{2}>u^{2}$. The string tension is
\beq
\sigma_{\rm V}=\lim_{v^{2}-u^{2}\rightarrow \infty}\;\frac{E_{\rm string}}{v^{2}-u^{2}} \;,
\nonumber
\eeq
where $E_{\rm string}$ is the lowest possible energy of the Hamiltonian
$H$ projected on the subspace of states with exactly one FZ particle for layers
with $x^{2}\ge u^{2}$ and $x^{2}<v^{2}$ and no FZ particles in any other
layer. To leading order $\sigma_{\rm V}=m/a$, where $m=m_{1}$ (for SU($2$)
there is only one mass). The projection of the Hamiltonian on this subspace is
\beq
H_{\rm proj}=\sum_{x^{2}=u^{2}}^{v^{2}-a}\sum_{k=1}^{4}\left\{m+\int \frac{dp}{2\pi}\;\frac{p^{2}}{2m}
{\mathfrak A}(p,x^{2})_{k}^{\dagger}
{\mathfrak A}(p,x^{2})_{k}\right\} +H_{1} \;, \label{primitive-effective-string}
\eeq
where ${\mathfrak A}(p,x^{2})_{k}$, ${\mathfrak A}(p,x^{2})_{k}^{\dagger}$ are the Fadeev-Zamolodchikov
destruction and creation operators (the field operator of the FZ particles), respectively,
with $x^{1}$-momentum $p$, the index $k=1,\dots, 4$ denotes the particle species
(the Hamiltonian is invariant under rotations in ${\rm O}(4)={\rm SU}(2)\! \times\! {\rm SU}(2)$)
where $H_{1}$ is given by (\ref{continuum-local}), as before. We are making
a nonrelativistic approximation. This approximation should be valid, provided
$(g_{0}^{\prime})^{2} \ll ma$ and we consider the lowest-lying states.

Particle states are produced on the vacuum by the application of FZ operators, e.g.
a one particle state with momentum $p$ and species index $k$ is
\beq
\left\vert p, k \right>={\mathfrak A}^{\dagger}(p)_{k}\vert 0>\;, \nonumber
\eeq
where the index $x^{2}$ is suppressed. In a theory of relativistic particles, these
states are normalized according to the rule
\beq
\left< p^{\prime}, k^{\prime}\right\vert p, k \left> \right.
=\frac{1}{\sqrt{p^{2}+m^{2}}}\delta_{k^{\prime} k }\delta(p^{\prime}-p)\;. \nonumber
\eeq

To find the spectrum of $H_{\rm proj}$, we need the matrix elements
\beq
\left< z_{1},k_{1}\right\vert j_{0}^{\rm L,R}(y) \left\vert z_{2},k_{2} 
\right>
=\left< z_{1}-y,k_{1}\right\vert j_{0}^{\rm L,R}(0) \left\vert z_{2}-y,k_{2} 
\right>\;, \nonumber
\eeq
where, for now, we have dropped the index $x^{2}$ and
where the particle states are given by 
$\left\vert z,k 
\right> ={\mathfrak A}(z)_{k}\left\vert 0 \right>$, $\left\vert 0 \right>$ being
the true vacuum of the ${\rm SU}(2)\times {\rm SU}(2)$ sigma model.

The matrix elements of currents may be written terms of momentum-space
eigenstates
by Fourier transformation:
\beq
\left< z_{1},k_{1}\right\vert j_{0}^{\rm L,R}(y) \left\vert z_{2},k_{2} 
\right>
&=&\int \frac{dp_{1}}{2\pi} \frac{1}{\sqrt{2E_{1}}}
\int \frac{dp_{2}}{2\pi} \frac{1}{\sqrt{2E_{1}}}
\nonumber \\
&\times& e^{-{\rm i}p_{1}(z_{1}-y)+{\rm i}p_{2}(z_{2}-y)}
\left< p_{1},k_{1}\right\vert j_{0}^{\rm L,R}(0) \left\vert p_{2},k_{2} 
\right>\;,
\label{Fourier-transformation}
\eeq
where $E_{1,2}={\sqrt{p_{1,2}^{2}+m^{2}}}$.
The momentum-space matrix elements have the exact
expression
\beq
\left< p_{1},k_{1}\right\vert j_{0}^{\rm L,R}(0) \left\vert p_{2},k_{2} 
\right>&=&
\frac{\rm i}{\sqrt 2} \left( \delta_{k_{1} 4}\delta_{k_{2}b}  -
\delta_{k_{2} 4}\delta_{k_{1}b}  \pm \epsilon_{b k_{1} k_{2} }\right) 
\nonumber \\
&\times&
(p_{1}+p_{2})F(\theta_{1}-\theta_{2}+{\rm i}\pi) , \label{useful-form-factor}
\eeq
where the plus or minus sign corresponds to the left-handed ($L$) or right-handed ($R$) current, respectively, the rapidities $\theta_{1,2}$ are defined by
$m\sinh\theta_{1,2}=p_{1,2}$, 
and
\beq
F(\theta)\!\!&\!\!=\!\!&\!\!\exp 2\int_{0}^{\infty} \frac{d\xi}{\xi}\, \frac{e^{-\xi}-1}{e^{\xi}+1}\,
\frac{\sin^{2} \frac{\xi(\pi{\rm i}-\theta)}{2\pi}}{\sinh \xi}  \nonumber \\
\!\!&\!\!=  \!\!&\!\! \exp -\int_{0}^{\infty} \frac{d\xi}{\xi}\, \frac{e^{-\xi}}{\cosh^{2}\frac{\xi}{2}}\,
\sin^{2} \frac{\xi(\pi{\rm i}-\theta)}{2\pi}. \nonumber
\eeq 
Note that the Kronecker deltas in (\ref{useful-form-factor}) are automatically zero
if an index takes the value $4$. This expression is the result of Karowski and Weisz \cite{KarowskiWeisz}
for the ${\rm O}(4)\simeq {\rm SU}(2)\times {\rm SU}(2)$ sigma-model form 
factors, after applying crossing \cite{PhysRevD74}.

The only difference between the free-field-theory matrix elements and (\ref{Fourier-transformation}), 
(\ref{useful-form-factor}) is the presence of the factor $F(\theta_{1}-\theta_{2}+{\rm i}\pi)$. The
physical interpretation of this factor is that the color of an FZ particle is not point-like, but
smeared over a region of size $m^{-1}$. This smearing will be made more
explicit in the discussion below.

Since the mass of the FZ particles is assumed large compared to $(g_{0}^{\prime})^{2}/a$,
we assume that in the frame where the sources
are static, these particles move slowly. We can therefore make the approximation
that $p_{1}$ and $p_{2}$
in the Fourier
transform in (\ref{Fourier-transformation}) are small compared to $m$. The result is
\beq
2^{-1/2}(p_{1}^{2}+\!\!&\!\!m^{2}\!\!\!&\!\!\!)^{-1/4} 2^{-1/2}(p_{2}^{2}+m^{2})^{-1/4}
 \left< p_{1},k_{1}\right\vert j_{0}^{\rm L,R}(0) \left\vert p_{2},k_{2} 
\right>   \nonumber \\
&=&\frac{\rm i}{\sqrt 2} \left( \delta_{k_{1} 4}\delta_{k_{2}b}  -
\delta_{k_{2} 4}\delta_{k_{1}b}  \pm \epsilon_{b k_{1} k_{2} }\right) 
\exp{-\frac{A}{m^{2}} (p_{1}-p_{2})^{2}  }\;, \label{Taylor-exp}
\eeq
where the positive constant $A$ is
\beq
A=\frac{1}{4\pi^{2}}\int_{0}^{\infty} d\xi \, \,\frac{\xi e^{-\xi}}{\cosh^{2}\frac{\xi}{2}}
=\frac{1}{12}-\frac{\ln 2}{\pi^{2}} =0.1310284\;.\nonumber
\eeq
It is convenient that, to leading order, all the momentum dependence
is in the exponent of (\ref{Taylor-exp}). This result just means that the color
distribution of an FZ particle is Gaussian. Inserting (\ref{Taylor-exp})
into (\ref{Fourier-transformation}) yields
\beq
\left< z_{1},k_{1}\right\vert j_{0}^{\rm L,R}(y) \left\vert z_{2},k_{2} 
\right>
={\sqrt {\frac{m^{2}}{4\pi A}}}(\sigma^{\rm L,R}_{b})_{kl}
\exp\left[-\frac{m^{2}}{4A}\left(\frac{z_{1}+z_{2}}{2}-y\right)^{2}\right] 
\delta(z_{1}-z_{2})\;. \label{smeared-charge}
\eeq
where the ``spin" operators are
\beq
(\sigma^{\rm L,R}_{b})_{kl}={\rm i}\left(\delta_{k 4}\delta_{lb}  -
\delta_{l 4}\delta_{kb}  \pm \epsilon_{b k l }\right) \;. \nonumber
\eeq
These operators are generators of independent spin-1/2
representations of color-SU($2$). Specifically, 
\beq
[\sigma^{\rm L,R}_{b},\sigma^{\rm L,R}_{c}]=2{\rm i}\epsilon_{bcd}\sigma^{\rm L,R}_{d}\;,\;\;
[\sigma^{\rm L}_{b},\sigma^{\rm R}_{c}]=0\;,\;\;
\sum_{b}(\sigma^{\rm L,R}_{b})^{2}=3\;.
\nonumber
\eeq

\section{The string Hamiltonian}
\setcounter{equation}{0}
\renewcommand{\theequation}{3.\arabic{equation}}

Next we use the smeared color-charge density (\ref{smeared-charge}) 
to write down the effective Hamiltonian of the string. We write $z=z(x^{2})$
for each value of $x^{2}$. From the interaction Hamiltonian (\ref{continuum-local}),
and the kinetic term in (\ref{primitive-effective-string}), this is
\beq
H_{\rm string}=
\frac{m}{a}(v^{2}-u^{2})-\frac{1}{2m}\sum_{x^{2}=u^{2}}^{v^{2}-a}\frac{\partial^{2}}{\partial z(x^{2})^{2}}
+V_{\rm bulk}+V_{\rm ends}\;, \nonumber
\eeq
where
\beq
V_{\rm bulk}&=&-\frac{m^{2}}{4\pi A}\frac{(g_{0}^{\prime})^{2}}{4g_{0}^{4}a^{2}}
\sum_{x^{2}=u^{2}+a}^{v^{2}-a} \int dx^{1} dy^{1} \vert x^{1}-y^{1} \vert \nonumber \\
&\times& \left\{ e^{-\frac{m^{2}}{4A}[z(x^{2})-x^{1}]^{2}} \sigma^{\rm L}(x^{2})_{b}
-e^{-\frac{m^{2}}{4A}[z(x^{2}-a)-x^{1}]^{2}} \sigma^{\rm R}(x^{2}-a)_{b}\right\}  \nonumber \\
&\times& \left\{ e^{-\frac{m^{2}}{4A}[z(x^{2})-y^{1}]^{2}} \sigma^{\rm L}(x^{2})_{b}
-e^{-\frac{m^{2}}{4A}[z(x^{2}-a)-y^{1}]^{2}} \sigma^{\rm R}(x^{2}-a)_{b}\right\}\;, 
\label{bulk-string-potential}
\eeq
and
\beq
V_{\rm ends}&=&-\frac{(g_{0}^{\prime})^{2}}{4g_{0}^{4}a^{2}}
\int dx^{1} dy^{1} \vert x^{1}-y^{1} \vert
\left\{{\sqrt{\frac{m^{2}}{2\pi A}}}
e^{-\frac{m^{2}}{4A}[z(u^{2})-x^{1}]^{2}} \sigma^{\rm L}(u^{2})_{b}
+\delta(x^{1}-u^{1}){\bar {\rm q}}_{b} \right\} \nonumber \\
&\times& \left\{ {\sqrt{\frac{m^{2}}{2\pi A}}}
e^{-\frac{m^{2}}{4A}[z(u^{2})-y^{1}]^{2}} \sigma^{\rm L}(u^{2})_{b}
+\delta(y^{1}-u^{1}){\bar {\rm q}}_{b} \right\} \nonumber \\
\!\!&\!\!-\!\!&\!\! \frac{(g_{0}^{\prime})^{2}}{4g_{0}^{4}a^{2}}
\int dx^{1} dy^{1} \vert x^{1}-y^{1} \vert
\left\{ {\sqrt{\frac{m^{2}}{2\pi A}}}e^{-\frac{m^{2}}{4A}[z(v^{2})-x^{1}]^{2}} \sigma^{\rm R}(v^{2}-a)_{b}
+\delta(x^{1}-v^{1}){\rm q}_{b} \right\} \nonumber \\
&\times& \left\{ {\sqrt{\frac{m^{2}}{2\pi A}}}
e^{-\frac{m^{2}}{4A}[z(v^{2})-y^{1}]^{2}} \sigma^{\rm R}(v^{2}-a)_{b}
+\delta(y^{1}-v^{1}){\rm q}_{b} \right\} \;. \label{end-string-potential}
\eeq

We need to apply the constraint (\ref{physical}) to states. This becomes
\beq
&\int \!\!&\!\!\! dx^{1} \left\{
-{\sqrt{\frac{m^{2}}{2\pi A}}} e^{-{\frac{m^{2}}{4A}[z(z^{2})-x^{1}]^{2}}}\sigma^{\rm L}(x^{2})_{b} \right.
\nonumber \\
&+& \left.{\sqrt{\frac{m^{2}}{2\pi A}}} e^{-{\frac{m^{2}}{4A}[z(z^{2}-a)-x^{1}]^{2}}}\sigma^{\rm R}(x^{2}-a)_{b}
\right\}
\Psi=0 \;, \nonumber
\eeq
for $x^{2}=u^{2}+a,\dots, v^{2}-a$, and
\beq
\int \!\!&\!\!dx^{1}\!\!&\!\! {\sqrt{\frac{m^{2}}{2\pi A}}}
\left\{
e^{-{\frac{m^{2}}{4A}[z(u^{2})-x^{1}]^{2}}}\sigma^{\rm L}(u^{2})_{b} -{\bar q}_{b}\delta(x^{1}-u^{1})
\right\}\Psi=0\;, \nonumber \\
\int \!\!&\!\!dx^{1}\!\!&\!\! {\sqrt{\frac{m^{2}}{2\pi A}}}
\left\{
e^{-{\frac{m^{2}}{4A}[z(v^{2}-a)-x^{1}]^{2}}}\sigma^{\rm L}(v^{2}-a)_{b} +q_{b}\delta(x^{1}-v^{1})
\right\}\Psi=0\;, \nonumber
\eeq
at the ends. These constraints simply reduce to the identification of $\sigma^{\rm L}(x^{2})_{b}$
with $\sigma^{\rm R}(x^{2}-a)_{b}$, for $x^{2}=u^{2}+a,\dots, v^{2}-a$, with
$\sigma^{\rm L}(u^{2})_{b}/\sqrt{2}$ with ${\bar q}_{b}$ and
$\sigma^{\rm R}(v^{2}-a)_{b}/\sqrt{2}$ with $-q_{b}$. In this way, the color degrees of
freedom are completely eliminated from (\ref{bulk-string-potential})
and (\ref{end-string-potential}).

There are integrals remaining to be done in 
(\ref{bulk-string-potential}), (\ref{end-string-potential}). One of these
is straightforward:
\beq
\int dx^{1} dy^{1} \vert x^{1}-y^{1} \vert e^{-\frac{m^{2}}{4A}[(x^{1})^{2}+(y^{1})^{2}]}
=\frac{4{\sqrt{2\pi}} A^{3/2}}{m^{3}}\;. \nonumber
\eeq
We write another integral we need as
\beq
\int dx^{1} dy^{1} \vert x^{1}-y^{1} \vert e^{-\frac{m^{2}}{4A}[(x^{1}+r)^{2}+(y^{1})^{2}]}
=\frac{4{\sqrt{2\pi}} A^{3/2}}{m^{3}}P(r) \;. \nonumber
\eeq
The third and final integral we need (simplifying the Hamiltonian near the
endpoints of the the string) is
\beq
\int d x^{1} \vert x^{1}-u^{1} \vert e^{-\frac{m^{2}}{4A}[x^{1}-z(u^{2})]^{2}}
=\frac{2A}{m^{2}}P[{\sqrt 2}z(u^{2})-{\sqrt 2}u^{1}] \;. \nonumber
\eeq
The function $P(r)$ cannot be evaluated exactly, but for small or large $r$ has the 
limiting forms
\beq
P(r)=\left\{\begin{array}{cc} 
1+\frac{m^{2}r^{2}}{4A}
\;,\;\;&  r\ll m^{-1}       \\ {\sqrt{\frac{\pi}{2A}}}m\vert r\vert \;, \;\;
& r\gg m^{-1}
\end{array}
\right.\;, \label{limiting-forms}
\eeq
respectively. We note
that the first of these forms can be derived from
the power series:
\beq
P(r)=1+\frac{m^{2}r^{2}}{4A}\sum_{n=0}^{\infty}\frac{(-1)^{n}}{(n+1)!(2n+1)}
\left( \frac{m^{2}r^{2}}{A}\right)^{n}\;. \nonumber
\eeq
The small-$r$ expression in (\ref{limiting-forms}) is due to
the softening of the linear potential in the horizontal direction from color smearing. At
large $r$, this smearing has no effect and the potential is linear.

Our result for the string Hamiltonian
is
\beq
H_{\rm string}\!\!&\!\!=\!\!&\!\!
\frac{m}{a}(v^{2}-u^{2})-\frac{1}{2m}\sum_{x^{2}=u^{2}}^{v^{2}-a}\frac{\partial^{2}}{z(x^{2})^{2}}
\nonumber \\
\!\!&\!\!-\!\!&\!\!\frac{3(g_{0}^{\prime})^{2}}{2g_{0}^{4}ma^{2}}{\sqrt{\frac{A}{2\pi}}}
\sum_{x^{2}=u^{2}+a}^{v^{2}-a} 
\left\{ 1-P[z(x^{2})-z(x^{2}-a)]
\right\}
\nonumber \\
\!\!&\!\!-\!\!&\!\! \frac{3(g_{0}^{\prime})^{2}}{2g_{0}^{4}ma^{2}}{\sqrt{\frac{A}{2\pi}}}
\left(1+ P\{{\sqrt{2}} [z(u^{2})-u^{1}]\} +P\{{\sqrt{2}} [z(v^{2}-a)-v^{1}]\} 
\right).
\label{final-string-Hamiltonian}
\eeq

\section{The static potential between sources}
\setcounter{equation}{0}
\renewcommand{\theequation}{4.\arabic{equation}}

Our result (\ref{final-string-Hamiltonian}) is simply a 
transversely-oscillating discretized Bosonic string. The only
unusual feature is that the potential energy becomes linear for large transverse
gradients. For small transverse gradients, however, the Hamiltonian
(\ref{final-string-Hamiltonian}) is quite conventional, since (\ref{limiting-forms})
yields a quadratic potential. We emphasize that this fortunate circumstance
is due entirely to the smearing of color of the FZ particles.
To determine the potential between static sources, we must find
the ground-state energy of (\ref{final-string-Hamiltonian}). This
is feasible because of the quadratic nature of the potential for small
gradients. In the small-gradient approximation
that $\vert z(x^{2})-z(x^{2}-a)\vert$, for $u^{2}<x^{2}<v^{2}$, 
$\vert z(u^{2})-u^{1}\vert$, and $\vert z(v^{2})-v^{1}\vert$ are
all much smaller than $m^{-1}$, the string Hamiltonian (\ref{final-string-Hamiltonian})
becomes
\beq
H_{\rm string}\!\!&\!\!=\!\!&\!\! \frac{3(g_{0}^{\prime})^{2}}{2g_{0}^{4}ma^{2}}{\sqrt\frac{A}{2\pi}}
+\frac{m}{a}(v^{2}-u^{2})
-\frac{1}{2m}\sum_{x^{2}=u^{2}}^{v^{2}-a}\frac{\partial^{2}}{\partial z(x^{2})^{2}}
\nonumber \\
\!\!&\!\!+\!\!&\!\! \frac{3(g_{0}^{\prime})^{2}}{8g_{0}^{4}ma^{2}}{\sqrt\frac{1}{2\pi A}}
\sum_{x^{2}=u^{2}+a}^{v^{2}-a} \left[
z(x^{2})-z(x^{2}-a)
\right]^{2}
\nonumber \\
\!\!&+\!\!&\!\! \frac{3(g_{0}^{\prime})^{2}}{4g_{0}^{4}ma^{2}}{\sqrt\frac{1}{2\pi A}}
\left\{
[z(u^{2})-u^{1}]^{2}+[z(v^{2}-a)-v^{1}]^{2}
\right\}\;. \label{quadratic-approx}
\eeq
Let us now drop the first, constant term in (\ref{quadratic-approx}) and denote $v^{2}-u^{2}$
by $L$.

The analysis of (\ref{quadratic-approx}) is straightforward. We drop the first term, which
has no physical significance. The potential in (\ref{quadratic-approx}) is diagonalized
by means of normal modes $w^{q}$, which have components:
\beq
(w^{q})_{k}=C^{q}\sin\left[\frac{\pi q}{Q}(k-\frac{1}{2})+\frac{\pi}{2}\right] \;, \nonumber
\eeq
where $k=(x^{2}-u^{2})/a$, $Q=(v^{2}-u^{2})/a=L/a$, $k,q=1,2,\dots,Q$ and
$C^{q}$ is a constant of normalization. If
we set $u^{1}=v^{1}$, then the Hamiltonian becomes a set of $Q$ simple harmonic
oscillators. The ground-state energy of $H_{\rm string}$ is
\beq
E_{0}=\frac{m}{a}L-\frac{{\sqrt 3}g_{0}^{\prime}}{g_{0}^{2}a}
\left(\frac{1}{2\pi A}
\right)^{1/4} \sum_{q=0}^{Q}\sin{\frac{\pi q}{2Q}}\;, \label{gs-energy}
\eeq
where all constant terms have been dropped. We apply the Euler summation
formula
\beq
\sum_{q=0}^{Q}F\left( \frac{q}{Q}\right)&=&Q\int_{0}^{1}dx \;F(x)-\frac{1}{2}\left[ F(1)-F(0)\right]
\nonumber \\
&+&\frac{1}{12Q}\left[ F^{\prime}(1)-F^{\prime}(0)\right] +{\mathcal O}\left(\frac{1}{Q^{2}}\right)
\;, \nonumber
\eeq
to 
(\ref{gs-energy}), and dropping constant terms once more, obtain the static
quark-antiquark potential
\beq
V(L)=E_{0}=\left[
\frac{m}{a}-\frac{2{\sqrt 3}}{\pi}\frac{g_{0}^{\prime}}{g_{0}^{2}a^{2}}
\right] L
-\frac{\pi{\sqrt 3}}{24} \frac{g_{0}^{\prime}}{g_{0}^{2}}\left(\frac{1}{2\pi A}
\right)^{1/4}\frac{1}{L}+{\mathcal O}\left( \frac{1}{L^{2}}\right)\;, \label{static-potential}
\eeq
which is our final result. Notice that in (\ref{static-potential}) there is a correction to the
string tension of order $g_{0}^{\prime}$, 
namely
\beq
\sigma_{\rm V}=\frac{m}{a}-\frac{2{\sqrt 3}}{\pi}\frac{g_{0}^{\prime}}{g_{0}^{2}a^{2}}\;.
\nonumber
\eeq
There is also a new term present in the potential proportional to
$1/L$. This term does not have the standard universal coefficient
\cite{LSW}, but instead is proportional to $g_{0}^{\prime}$.

\section{Some remarks on the isotropic case}
\setcounter{equation}{0}
\renewcommand{\theequation}{5.\arabic{equation}}

The picture
of confinement in the anisotropic theory is sufficiently compelling that we believe
the behavior of the standard rotationally-invariant theory is fundamentally similar. The 
necessity of the inequality (\ref{relative-scales}) shows 
that the rotationally-invariant theory is not easily accessible by the methods discussed
in this section. We argued that applying an anisotropic renormalization group causes
a theory for which $g_{0}^{\prime}\approx g_{0}$ to flow to
$g_{0}^{\prime}\ll g_{0}$ in the infrared \cite{PhysRevD75-2}. This infrared form of the
theory is essentially just 
a nonrelativistic approximation for the isotropic theory. A 
theory with a mass gap has a nonrelativistic limit
(the classical Yang-Mills theory, which is massless, has no such limit). Consider the Yang-Mills
action in $2+1$ dimensions with the speed of light included explicitly:
\begin{eqnarray}
{\mathcal S}=\frac{1}{c}\int d^{2}x \,dt \;{\rm Tr}\left[\frac{1}{2e^{2}}\sum_{i=1}^{2}(F_{0i})^{2}
-\frac{c^{2}}{2e^{2}}(F_{12})^{2}\right]\;, \nonumber
\end{eqnarray}
where $e$ is the continuum coupling. Suppose we 
Wick rotate this action to Euclidean space by $x^{0}\rightarrow {\rm i}x^{0}$, rotate 
so that $F_{12}\rightarrow F_{01}$, and finally
Wick rotate back. By identifying $g_{0}=e/\sqrt{a}$ and $g_{0}^{\prime}=e/({\sqrt a}c^{2})$, where
$a$ is a cut-off with units of centimeters,  
and taking $c\gg1$, our naive result is just 
the anisotropic model
discussed in this and previous papers. Certain observables in the anisotropic gauge theory
can now be identified with observables in standard Yang-Mills theory, with a caveat. The
caveat is that the mass scale is given by (\ref{mass-spectrum}) rather than being
proportional to the continuum coupling (this is because the justification for this procedure
relies on the anisotropic renormalization group argument given above). 

After the rotations described above, the string tension would 
be given by the space-like Wilson loop. By $(1+1)$-dimensional Lorentz invariance,
that is exactly the vertical string tension, studied in this paper. Now
the ratio of the string tension (which is $\sigma_{\rm V}$) to the square of the mass gap $M$
of the isotropic theory
can be obtained by examining correlation functions 
\beq
\left< j^{\rm L,R}_{\mu}(x^{1},x^{2})j^{\rm L,R}_{\nu}(x^{1},x^{2}+T)\right>
\sim \exp-Mc^{2}T\;,
\nonumber
\eeq
for large $T$. This would be
the first calculation of this ratio which is neither numerical, nor
relying on strong-coupling expansions. If this idea can be made to work, the
term proportional to $1/L$ in the potential (\ref{static-potential}) should have the universal coefficient
of reference \cite{LSW}.

\section{Conclusions}

To summarize, we have determined the potential between static
sources, separated in the $x^{2}$-direction in $(2+1)$-dimensional
SU($2$) Yang-Mills theory with two couplings $g_{0}$ and $g_{0}^{\prime}$. The
calculation, like those in 
\cite{PhysRevD75-1},
\cite{PhysRevD74},
\cite{PhysRevD75-2},
is done entirely in a weak-coupling approximation, in which $g_{0}^{\prime}$ is 
smaller than any power of $g_{0}$. The non-point-like nature of
the color charge of the fundamental
excitations of the principal-chiral sigma model is essential to understanding
the result. The physical string states are color singlets by virtue of Gauss's law. This
feature should also be the case for gauge group SU($N$); unfortunately, nothing
explicit can be done for $N>2$, as the generalization of (\ref{useful-form-factor})
is not known.

The composite-string Hamiltonian (\ref{quadratic-approx}), describing the electric flux between
a verti-\\
cally-separated quark-anti-quark pair, can be studied by several techniques, among
them numerical. One
could eventually imagine real-space renormalization-group  or numerical variational
methods applied to this
problem.

Using the exact S-matrix for FZ particles, the scattering problem of string states, either
mesonic, such as those we have considered here, or purely gluonic, can be studied. In particular, amplitudes at
large center of mass energies, {\em i.e.} Pomerons, in which 
gluonic processes dominate, are analytically accessible. Calculating the
scattering amplitude in this asymptotically-free version of $(2+1)$-dimensional
Yang-Mills theory may give some general insight into large-$s$ scattering.

In Section 5, we conjectured
that ratios of some quantities in the isotropic theory may be determined by 
those in our anisotropic model, through anisotropic renormalization flow. If this is the case, the string tension in the
isotropic theory is proportional to the vertical string tension, {\em i.e.} that studied
in this paper. For $N>2$, this would also mean that the $k$-string tensions in the isotropic
theory should be
proportional to 
$\sin\pi k/N$ - for this is true of the vertical $k$-string tensions of our model
\cite{PhysRevD75-1}.  This sine-law behavior  
was found in models of
strong-coupling QCD; in particular,
${\mathcal N}=2$ supersymmetric gauge theory softly broken to ${\mathcal N}=1$
\cite{DouglasShenker}, in M-theory 5-brane QCD \cite{HananyStrasslerZaffaroni},
and in the AdS/QCD scheme \cite{HerzogKlebanov}. The sine law 
was indicated in
one calculation \cite{DelDebbio2}, but most simulations
in four dimensions point to a result 
between the so-called Casimir law and the sine law \cite{DelDebbio1}, \cite{Lucini1}, 
\cite{Lucini2}. In $(2+1)$ dimensions, Bringoltz and Teper's recent
results indicate that the sine law does not
hold \cite{BringTep}. This would bode ill for the conjecture of 
Section 5, unless corrections to these string tensions
of order $g_{0}^{\prime}$ have significant $1/N$ dependence
(thus far, we can find results like (\ref{static-potential}) only for $N=2$). We 
hope that behavior of $k$-string tensions
will be settled soon, as more large-scale lattice simulations are carried out.

\section*{Acknowledgments}

We thank Alexei Tsvelik for a discussion about color smearing and
the string Hamiltonian in the small-gradient approximation.

This research was supported in
part by the National Science Foundation, under Grant No. PHY0653435
and by a grant from the PSC-CUNY.

\end{document}